\documentclass[a4paper,fleqn,usenatbib]{mnras}

\usepackage{ae,aecompl}

\usepackage{graphicx}	
\usepackage{amsmath}	
\usepackage{amssymb}	
\usepackage{rotating}
\usepackage{listings}
\usepackage{threeparttable}
\usepackage{pdflscape}

\title[The INTEGRAL/IBIS AGN catalogue: an update ] 
{\emph {The INTEGRAL/IBIS} AGN catalogue: an update}
\author[A. Malizia et al. ]
{A. Malizia,$^1$\thanks{E-mail address: \texttt{malizia@iasfbo.inaf.it}}, 
  R. Landi,$^1$  M. Molina,$^1$ L. Bassani,$^1$ A. Bazzano,$^2$ A. J. Bird,$^3$  \newauthor   P. Ubertini$^2$ \\
$^1$ INAF/IASF-Bologna, Via P. Gobetti 101, I-40129 Bologna, Italy \\
$^2$ INAF/IASF-Roma, Via Fosso del Cavaliere 100, I-00133, Roma, Italy \\
$^3$ School of Physics and Astronomy, University of Southampton, SO17 1BJ, Southampton, UK 
}

\date{Accepted XXX. Received YYY; in original form ZZZ}

\pubyear{2016}

\begin{document}
\label{firstpage}
\pagerange{\pageref{firstpage}--\pageref{lastpage}}
\maketitle

\begin{abstract}
In the most recent  \emph{IBIS} survey based on  observations performed during the first 1000 orbits of \emph{INTEGRAL}, are listed 363  high energy emitters firmly associated with AGN, 107 of which 
are reported here for the first time.
We have used X-ray data to image the \emph{IBIS} 90\% error circle of all the AGN in the sample of 107, in order to obtain the correct X-ray counterparts, locate them with arcsec
accuracy and therefore pinpoint the correct optical counterparts. This procedure has led to the optical and spectral characterization of the entire sample. 
This new set consists of 34 broad line or type 1 AGN, 47 narrow line or type 2 AGN, 18 Blazars and 8 sources of unknown class. 
These 8 sources have been associated with AGN from their positional coincidence with 2MASX/Radio/X-ray sources.
Seven high energy emitters have been included since they are considered to be good AGN candidates.
Spectral analysis has been already performed on 55 objects and the results from the most recent and/or best statistical measurements have been collected.
For the remaining 52 sources, we report  the spectral analysis for the first time in this work. We have been able to obtain the full X-ray coverage of the sample making use of data from \emph{Swift/XRT}, \emph{XMM-Newton}
and \emph{NuSTAR}. In addition to the spectral characterization of the entire sample, this analysis has enabled us to identify peculiar sources and by comparing different datasets, highlight flux variability in the 2-10 keV and
20-40 keV bands.
\end{abstract}




\begin{figure*}
\includegraphics[width=\textwidth,width=8cm,angle=90]{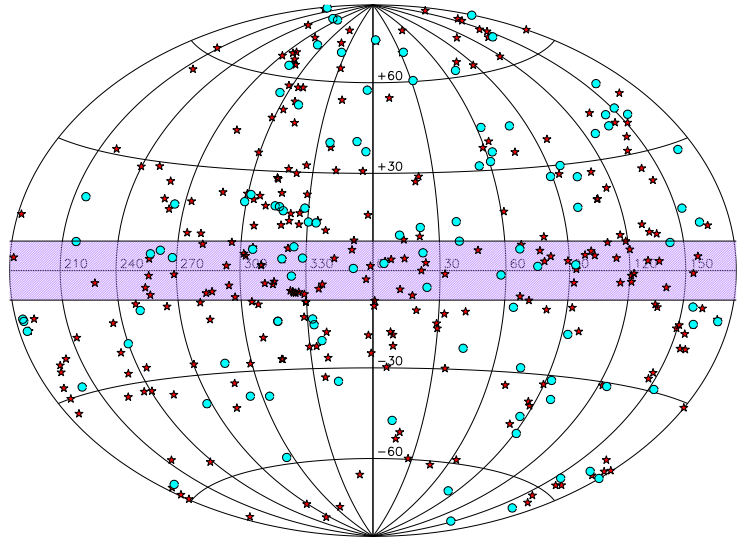}
\caption{All the AGN so far detected by \emph{INTEGRAL/IBIS} plotted on the sky. The stars represent the 107 new active galaxies studied in this work and firstly reported as \emph{INTEGRAL} AGN in the last \emph{IBIS} survey \citep{bird:2016}, 
while the circles are the AGN detected in previous surveys by \emph{INTEGRAL/IBIS}.}
\label{fig1}
\end{figure*}

\section{Introduction}
In the last decade both \emph{INTEGRAL/IBIS} \citep{ubertini:2003} and \emph{Swift/BAT}  \citep{barthelmy:2005}, having good sensitivity and a wide-field sky coverage, were able to make progress in the high energy domain (20-200 keV).
In particular they have provided a great improvement in the knowledge of the high energy extragalactic sky by detecting almost 1000 (mostly local) 
AGN at energies above 15 keV (\citealt{malizia:2012}, \citealt{baumgartner:2013}, this work).

Recently a new all sky \emph{INTEGRAL} catalogue has been published \citep{bird:2016}  containing 939 hard X-ray selected sources of which 363  are definitely associated with AGN.
Of these 107  have been reported for the first time as \emph{INTEGRAL/IBIS} detections and another 7 objects are considered to be good candidate active galaxies. 
Thus extragalactic objects represent a significant fraction of the entire catalogue (about 40\%) and provide a well defined sample of sources which can be used for statistical and population studies of AGN  in many astrophysical contexts.
In particular, the hard X-ray band is ideal to obtain information on the  intrinsic column density of active galaxies (and hence its relation with other source parameters) since this waveband is the least affected by absorption. 
By comparing the hard X-ray emission with that of other wavebands, the \emph{INTEGRAL} sample can be used for extensive correlation studies; it is also ideal for  spectral analysis since this band is unique for studying spectral features 
such as the Compton reflection bump and the high energy cut-off.

Figure 1 shows the distribution of all AGN so far detected by \emph{INTEGRAL} plotted on the sky; the stars represent the 107 new galaxies  reported in this work, 
while the  circles represent the AGN detected in the previous \emph{IBIS} catalogues.  The horizontal strip between $\pm$10$^{\circ}$ in Galactic latitude is shown to highlight the so called Zone of Avoidance in the Galactic Plane 
where  INTEGRAL plays a key role in detecting new absorbed objects and in particular AGN. In fact, due to observational strategy, \emph{INTEGRAL} 
is more effective in finding hard X-ray emitting sources along the Galactic Plane and  along the directions of the spiral arms while \emph{Swift/BAT} is more effective at higher Galactic latitudes.
Indeed among  the 72 AGN detected by \emph{INTEGRAL/IBIS} and not by \emph{Swift/BAT}, almost  50\% (37 objects) are in the Galactic plane while the remaining sources are either detected as variable in the \emph{IBIS} survey or are located in regions
of the sky where \emph{BAT} has less exposure. 

A large fraction of the  \emph{INTEGRAL}  AGN present in the new survey of \cite{bird:2016} have already been fully characterized in terms of optical and X-ray properties in \cite{malizia:2012}, while the 107 new entries 
are presented and characterized in terms of their optical and X-ray properties in this paper where a few interesting objects are also highlighted and discussed.

\section{INTEGRAL AGN update}
In this paper we present the X-ray and optical follow-up work on  107 new AGN recently detected by \emph{INTEGRAL}. 
Luckily, we have been able to obtain full X-ray coverage of the entire sample making use of data from  \emph{Swift/XRT}, \emph{Newton-XMM} and \emph{NuSTAR} 
archives or through \emph{Swift/XRT} follow up observations triggered by us. 

2-10 keV  data are important for two main reasons: 1) they provide a unique tool to associate the high energy source with a single/multiple X-ray counterpart, getting a better position and 2) to characterize the high energy 
source at low energies, for example by estimating the spectral shape, the intrinsic absorption and  the 2-10 keV flux/luminosity.

Association is a fundamental  step in the analysis as it allows us to pinpoint the correct optical counterpart, locate it with arcsec accuracy and therefore provide a way to classify the source through optical spectroscopy. 
We have used the X-ray data to image the \emph{IBIS} 90\% error circle of all AGN in the sample of 107 in order to determine the correct X-ray counterpart and in the majority of cases we found a single clear association.
However, there were several sources where the \emph{XRT} field of view contained  more than one X-ray detection, thus deserving a deeper investigation.
In these cases in order to pinpoint all  counterparts and select the most likely, we superimposed the  99\% high energy error circle on the 3--10 keV  X-ray image.
This allowed us to associate the hardest and the brightest X-ray source with the high energy emitter.
Generally secondary X-ray detections disappear above 3 keV or result in having a much lower signal to noise ratio with respect to the brightest source. This was the case for SWIFT J0444.1+2813,
SWIFT J1630.5+3925, IGR J18078+1123, IGR J18308+0928  and SWIFT J1933.9+3258 where at energies higher than 3 keV a secondary X-ray object is still detected but only at lower significance level.
We also note that in all but one case the X-ray counterpart was located well inside the \emph{INTEGRAL} positional uncertainty; the only exception is SWIFT J1436.8-1615
for which a single X-ray source, a QSO at redshift of 0.14454, was found just at the border of the \emph{IBIS} error circle.

As a final remark, we note that this accurate method of association, led us to discover also an incorrect   identification previously reported in the literature. This is the case of SWIFT J1238.6+0828 which has been associated with the galaxy VCC 1759
at z=0.0321 \citep{baumgartner:2013}. From the inspection of the \emph{XMM-pn} image there is no X-ray detection from VCC 1759 while there is 2-10 keV  emission from the galaxy cluster WHL J123841.5+092832  at z = 0.229900 and
from the Sey 2 galaxy 2MASX J12384342+0927362 at z= 0.0829. Both objects are well inside the \emph{IBIS} and \emph{BAT} error circles. 
As clearly shown in Figure 2, which displays the \emph{XMM-pn} 4.5-12 keV image, the most likely X-ray counterpart of SWIFT J1238.6+0928
is the Sey 2 galaxy  2MASX J12384342+0927362, since the cluster is no longer detected above 4.5 keV.

Prior to characterising  each source spectrum at low energies, we have also pinpointed the correct optical counterpart and searched the literature and relevant archives for its  appropriate classification. 
This has led to the optical characterization of almost all of the 107 new AGN. Optical classes have been mainly collected from NED (NASA Extragalactic database), 
SIMBAD and from the 13th edition of the  extragalactic catalogue of \cite{veron:2010}. 
Table A1 lists all the new 107 AGN\footnote {Table A1 lists 108 entries since IGR J16058-7253 is a galaxy pair with both AGN detected in the X-ray band} together with their alternative names, coordinates,\footnote{All the coordinates have been taken
from the NED archive} redshift, class (columns from 2 to 6) 
and X-ray spectral parameters (columns 7 to 11, see section 3). The \emph{INTEGRAL/IBIS} 20-100 keV fluxes extracted from \cite{bird:2016} are reported in column 12, while in the last column (13) the references for the X-ray spectral parameters are listed.
88 AGN/candidate AGN were detected for the first time in the hard X-ray band by \emph{INTEGRAL} and reported in \cite{bird:2016},  while the remaining 26 objects, labelled in the table A1 with the symbol $^{\clubsuit}$, 
were already present in previous \emph{IBIS} catalogues but only recently identified as active galactic nuclei and hence reported here as new \emph{INTEGRAL} AGN.
For 11 objects we have been able to secure optical follow up work and hence in these cases the source class is reported here for the first time while details of the optical spectra will be provided in a forthcoming paper by Masetti et al. (in
preparation); these 11 objects have been labelled in the table with the ($\S$) symbol.

In Table A1 there are 8 objects for which the optical class is still missing but their association with an AGN has been argued from different indicators  such as the
positional coincidence with a galaxy associated with a 2MASX extended object and/or with a radio source  from which X-ray emission has been detected. Their class in Table A1 has been labelled as {\it AGN}.
All but 2 of these  sources where already  known and extensively discussed in the literature (see their corresponding references).
The two exceptions are IGR J17111+3910 and PBC J1850.7-1658:  the first is positionally coincident with a \emph{Rosat} (1RXS J171105.6+390850) and a 2MASX source (2MASXJ17110531+3908488), 
while the second is associated with a bright radio source  (NVSS J185054-165543) of 390 mJy flux at 20 cm. Both of them have an X-ray counterpart which has been observed by \emph{Swift/XRT} and
whose data have been analysed and presented in this work for the first time.
For these 8 AGN and 4 objects classified as Blazars, redshift measurements are still  missing. 

Following our previous works \citep{malizia:2007, malizia:2012}, this new set of AGN can be divided in 34 broad line or type 1 AGN (Seyfert 1-1.5), 47 
narrow line or type 2 (type 1.8-2), 18 Blazars and 8 sources  of unknown class. KAZ 320 is the only Narrow Line Seyfert galaxy (NLS1) and is included in the type 1 AGN subset while the only
Liner (NGC 5100), being absorbed, has been considered as a type 2 object. 

\begin{figure}
\includegraphics[width=1.0\linewidth]{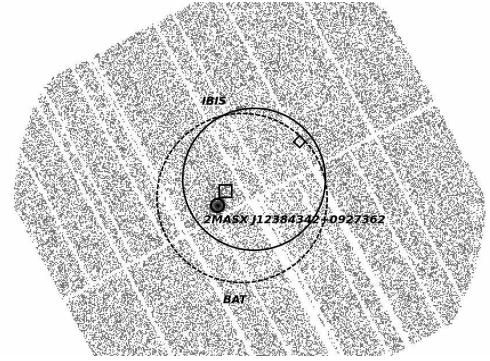}
\caption{\emph{XMM-pn} 4.5-10 keV image of the field around SWIFT J1238.6+0828. The solid and the dashed circles are the \emph{IBIS} and \emph{BAT} error circles respectively. It is clear that there is only one X-ray detection
corresponding to the Sey 2 galaxy 2MASX J12384342+0927362 at z= 0.0829. The box point shows the position of the galaxy cluster WHL J123841.5+092832 who does not emit any more at 4.5 keV, while the diamond point
corresponds to the position of the galaxy VCC 1759, previously associated with SWIFT J1238.6+0828. }
\label{fig2}
\end{figure}

\subsection{AGN candidates}
For the sake of completeness we have also added at the end of Table A1, 7 sources which  we consider to be  good AGN candidates.

IGR J02447+7046  has been optically identified as a Sey 1.2 galaxy by \cite{masetti:2013} and indeed, analyzing the \emph{XRT} data, a soft
X-ray source coincident with the optical position of this AGN has been marginally detected ($\sigma$ $\sim$ 4) although just outside the 90\% high energy error circle. 
In order to improve the detection statistics, we have requested and obtained  an extra \emph{XRT} observation. The addition of these data reveals the presence of another X-ray source well inside the \emph{IBIS} error box 
at a significance level of $\sim$ 4.4$\sigma$ and positionally coincident with the radio source NVSS J024443+704946 having  a 20cm flux of 204 mJy.
Since both sources weakly emit above 3 keV,  are compatible with the \emph{IBIS} error box and detected  at almost the same significance level, 
we assume that both contribute to the high energy emission. 
However,  since IGR J02447+7046 is detected in a 3.4 day outburst  in the IBIS survey \citep{bird:2016} it is more likely that the high energy emission is associated 
with a blazar type source that could possibly be linked to NVSS J025553+704946.

1RXS J112955.1-655542 has been considered as a candidate active galaxy since it is included in the WISE/2MASS/RASS AGN sample catalogue collected by \cite{edelson:2012} but it is the only object for which no X-ray information
is available. 

The remaining five objects have been proposed as AGN candidates in the literature before  and the relevant reference for each one is reported in column 12 of Table A1.

\section{X-ray data analysis}
For all the sample sources presented in this paper, we have collected the spectral properties in the contiguous 2-10 keV X-ray band in order to better characterize their 
nature and to search for spectral peculiarities.  While for 55 objects, X-ray spectral  analysis has already been performed and the results published,
for the remaining 52 sources (almost half of the sample) we here report  the X-ray spectral analysis for the first time.
For the 55 already studied objects, spectral parameters are taken from the most recent and/or best statistical significance measurement (the relevant reference is reported in Table A1 as well as the instrument used).
For all the remaining sources X-ray data have been retrieved from \emph{Swift/XRT}, \emph{Newton-XMM} and \emph{NuSTAR} public archives available before the end of 2015.
In cases where for a source we found observational data in more than one archive, the observation with the best statistics has been used for the data analysis and the results of the best fits are reported in Table A1. 

All sources have  been observed with the \emph{Swift/X-Ray Telescope (XRT)} with the exception of
PKS 0312-770, SWIFT J0709.3-1527  and SWIFT J1238.6+0928, for which \emph{XMM-Newton} observations were available. 
Furthermore, 7 of the 52  AGN (PKS 0637-752, SWIFT J0845.0-3531, SWIFT J1410.9-4229, MCG-01-40-00, IGR J18308+0928, 4C +21.55 and UGC 12049) have 
\emph{NuSTAR} data available although only in four sources these data are of better statistical quality than the \emph{XRT} ones in the same energy band.
In  columns 6 and 7 of Table A1 the net exposure  and the signal to noise ratios are listed; the exposures with no label correspond to the \emph{XRT} data while the  sources for which 
\emph{XMM} or \emph{NuSTAR} data have been used are labelled with "X" and "N" respectively.

All spectral fitting was performed in \texttt{XSPEC} v.12.8.2  using the solar abundances of \cite{anders:1989}.
Uncertainties are listed at the 90\% confidence level ($\Delta \chi^{2}$ = 2.71 for one interesting parameter).

\subsection{Swift/XRT data}
{\emph{XRT} data reduction was performed using the standard data pipeline package (\texttt{XRTPIPELINE} v. 0.13.2) in order to produce screened event files following the procedure described in \cite{landi:2010}.
Due to the pointing strategy of \emph{Swift}, short (a few ksec) repeated measurements are typically performed for each target, therefore to improve the statistical
quality of the data, for those sources with more than one pointing, we performed the spectral analysis of each observation individually and then of  the combined spectra. Indeed we found in most cases 
that  spectra from individual pointings were compatible with each other within the respective uncertainties, thus justifying a combined analysis. In case of spectral variability the single observations have
been considered and compared to each other (see following sections).

The \emph{XRT} spectral analysis of 44 sources  out of 52 with no published X-ray data have been performed in the best energy range depending on the detection significance of the pointed source. 
As evident from the values reported in column 7 of table A1, we do not always have sufficient statistics to investigate the spectral behaviuor in the 2-10 keV energy range in depth. Therefore in order to characterize the X-ray spectra 
we have employed a simple power law model absorbed by the Galactic column density and, if required, by intrinsic absorption. 
In  a number of cases (11 sources) it was necessary to fix the photon index at the canonical value of $\Gamma$=1.7 \citep{malizia:2014} in order to estimate the absorption and the 2-10 keV flux. 
Generally the \emph{XRT} exposures are not sufficient to detect the iron line at around 6 keV and therefore we could not investigate the presence of this important feature in the X-ray spectrum of AGN. 
This feature thought to be due to a fluorescence line from the K-shell of iron and produced when the nuclear continuum radiation is reprocessed by circumnuclear material, could allow us to investigate the 
site of reprocessing through its width and strength suggesting its origin
in the outer broad line region and/or in the molecular torus.\\
In the following sections a few peculiar sources found when analysing the \emph{XRT} data are discussed in more details.

\subsubsection{Galaxy pairs: ESO 328-36 and IGR J16058-7253}
ESO 328-36 is actually a pair of galaxies at z=0.02370  very close to each other (0.8 arcmin) of which only one (ESO 328-IG 036 NED01) is an AGN classified as a Sey 1.8. 
Indeed \emph{XRT}, which has an angular resolution of 18 arcsec and therefore is able to discriminate them, has detected only the Seyfert galaxy in the 2-10 keV band. 
Its X-ray spectrum is well fitted by a power law with $\Gamma$ = 1.77 with no intrinsic absorption and  a 2-10 keV flux of $\sim$ 7 $\times$ 10$^{-12}$ erg cm$^{-2}$ s$^{-1}$.
The lack of intrinsic absorption is at odds with the optical classification of the source and this discrepancy could indicate the presence of dust, possibly a bridge due to the galaxy merging, which obscures the 
broad lines in the optical band. 

Another interesting case is the "false" pair associated with the high energy emitter IGR J16058-7253 which has already been highlighted as a pair of sources both detected in the X-ray band by \cite{landi:2011a}.
Following the  optical follow up work it has been assessed that the two AGN, LEDA 259580 and LEDA 259433, which are 3.4 arcmin apart, have different redshifts, 0.090  and 0.069 respectively, making them a pair only due to perspective.
Of course with the \emph{IBIS} angular resolution of 12 arcmin, the high energy emission is a blending of the two galaxies and in this work we have re-analised  \emph{XRT} data in order to estimate the 20-100 keV flux of
each of them extrapolating from the X-ray spectrum. The 2-10 keV spectral parameters we obtained are in good agreement with \cite{landi:2011a} and from these we have estimated a 20-100 keV flux of 
 $\sim$ 10$^{-11}$ erg cm$^{-2}$ s$^{-1}$ for LEDA 259580 and $\sim$ 6.5 $\times$ 10$^{-12}$ erg cm$^{-2}$ s$^{-1}$ for LEDA 259433.

\subsubsection{IGR J17379-5957 = ESO 139-G012}
IGR J17379-5957 is  a new AGN from the last \emph{IBIS} catalogue although it has already been  reported  at high energies by \emph{Swift/BAT}.
It is associated with the galaxy ESO 139-G012 at z=0.01702,  classified as a Seyfert 2 by \cite{veron:2010}; however its optical spectrum does not
extend to [O III] and H${\beta}$ wavelengths and the indication of its AGN nature comes from the hint of a broadened H${\alpha}$ line \citep{marquez:2004}.
In the X-ray band it has been observed by \emph{Chandra} since ESO 139-G012 belongs to a set of AGN which have been correlated with the ultrahigh energy cosmic 
rays (UHECRs) observed by the \emph{Pierre Auger Collaboration} \citep{terrano:2012}.
\emph{Chandra} observation has been performed in order to estimate the source bolometric luminosity since only an upper limit to the X-ray flux was available until recently \citep{zaw:2009}.
\cite{terrano:2012} fitted the  \emph{Chandra} data assuming an absorbed power law which yielded  a hard photon index ($\Gamma$=0.72 $\pm$0.09),
a very low absorption N$_{H}$ $<$ 8.2 $\times$ 10$^{21}$ cm$^{-2}$ and a 2-10 keV flux of 4.7$^{+0.6}_{-1.1}$ $\times$ 10$^{-12}$ erg cm$^{-2}$ s$^{-1}$.
The hardness of the X-ray spectrum and the lack of absorption could mimic a Compton thick AGN and hence a possible wrong estimate of the source
bolometric luminosity. 

We checked the \emph{XRT} archive and found 9 observations of ESO 139-G012, of which 6 have been performed in 2008 and 3 in 2013.
We have first performed the spectral analysis of each single observation  except for the three performed in 2013 which were too short and near in time and therefore have been  summed 
in order to have enough statistics. A simple model consisting of a power law absorbed by the Galactic column density has been employed giving a good fit for each observation.
No intrinsic absorption is required by the data in agreement with the \emph{Chandra} results but we get a canonical photon index ($\Gamma$ $\sim$1.9) instead of the one found by \cite{terrano:2012}. 
In Table 1 we have reported the spectral parameters obtained for each dataset; it is evident  that while the spectral shape does not change, 
this source exhibits a quite strong flux variability on short timescales (days/months). Since the spectral shape is constant we summed together all the observations
getting a good quality spectrum ($\sigma$=62) which enabled us to better investigate  the spectral behaviour of IGR J17379-5957. Indeed from the total spectrum
soft excess emission below 2 keV is evident, together with  a hint of an iron line around 6.4 keV. Using the model {\texttt wa(mekal+po+zga)} in \texttt {XSPEC}, we find that our best fit 
($\chi^{2}$=171, 137 d.o.f.) provides  a gas  temperature of 0.19 keV and a photon index of  1.8$\pm$0.07. The iron line is required by the data at  $>$95\% confidence level, its energy is 6.45 keV 
with an equivalent width (EW) ranging from 126 to 636 eV. The lack of  intrinsic absorption and X-ray flux variability are 
at odds with the source classification, making IGR J17379-5957 an even more interesting object. 
Being a variable source, it would be better to obtain the optical class and the column density measurement through simultaneous observations. Furthermore, it is possible that the source is
an intermediate type 2 Sey (1.8-1.9) since a broad H${\alpha}$ line was detected previously. In these sources, optical flux variability can lead to a source 
misclassification \citep{trippe:2010}.

\begin{table}
\begin{center}
\caption{Swift-XRT observation of IGR J17379-5957 (ESO 139-G012)}
\begin{tabular}{llccc}
\hline
Date                &  Expo (s)   &    $\sigma$      & $\Gamma$           & F$_{2-10~keV}^{\dagger}$ \\
\hline\hline
05/04/2008     &  1940          &    24       & 2.02$^{+0.15}_{-0.10}$   & 7.28$^{+1.35}_{-0.90}$ \\
06/03/2008     &  2640          & 11.5       & 1.80$^{+0.35}_{-0.38}$   & 1.93$^{+1.02}_{-0.59}$ \\ 
07/10/2008     & 552             & 11          & 2.05$^{+0.30}_{-0.30}$   & 5.40$^{+1.79}_{-1.32}$ \\  
07/21/2008     & 4982           & 36          & 1.97$^{+0.08}_{-0.09}$   & 6.27$^{+0.46}_{-0.75}$ \\
07/29/2008     & 3339           & 26          & 1.97$^{+0.12}_{-0.11}$   & 4.68$^{+0.87}_{-0.44}$  \\
11/02/2008     & 2063           & 26          & 1.89$^{+0.12}_{-0.12}$   & 8.85$^{+0.91}_{-0.85}$  \\
Mar-Oct 2013 &    1831        & 13         & 1.83$^{+0.21}_{-0.21}$    & 2.98$^{+0.52}_{-0.50}$  \\     
\hline
total$^{\star}$ &  17350         &  62       & 1.80$^{+0.07}_{-0.07}$     & 6.02$^{+0.27}_{-0.22}$  \\
\hline
\end{tabular}
\end{center}
$\dagger$ in units of $\times$ 10$^{-12}$ erg cm$^{-2}$ s$^{-1}$;\\
$\star$ values referred to the best fit model (see text).
\end{table}

\subsection{XMM-Newton data}
For the four sources observed by \emph{XMM-Newton} (MRK 1152, PKS 0312-770, SWIFT J0709.3-1527 and SWIFT J1238.6+0928) only the \emph{pn} camera data have been considered.
The data were reprocessed using the \emph{XMM-Newton} Standard Analysis Software (SAS) version 14.0 employing the latest available calibration
files following the standard procedure illustrated in e.g. \cite{molina:2009}. The 0.3-10 keV energy range has been used to fit the data and
in most cases the presence of a soft excess and an iron line in addition to  a power law with Galactic and intrinsic absorption has been investigated.
Indeed a soft excess has been detected in MRK 1152 and PKS 0312-770 (kT =0.19 and 0.35 keV respectively) both classified as Sey 1; also in SWIFT J1238.6+0928
associated with the Sey 2 galaxy 2MASX J12384342+0927362 we found evidence of a soft excess,  but its nature deserves a deeper investigation due to the presence of the nearby cluster that could 
contaminate the source spectrum at soft energies (see section 2).

\subsubsection{SWIFT J0709.3-1527 = PKS 0706-15}
SWIFT J0709.3-1527 is a new AGN listed in the last \emph{IBIS} survey which has been associated with the radio source PKS 0706-15 and whose optical spectrum
has been recently acquired revealing it to be a BL Lac object (Masetti et al. in preparation).
Its spectral shape is consistent with a simple  power law having a photon index $\Gamma$=2.31$\pm$0.06 absorbed only by the Galactic column density; the fit does not require 
any spectral curvature intrinsic or due to local absorption as usually found in the X-ray spectra of BL Lac objects. 
The 2-10 keV flux is 1.36 $\times$ 10$^{-11}$ erg cm$^{-2}$ s$^{-1}$ while at high energy it is detected by \emph{IBIS} at 4.5$\sigma$ in the 20-40 keV band with a flux of 
6.8 $\times$ 10$^{-12}$ erg cm$^{-2}$ s$^{-1}$ and has only an upper 
limit in the 40-100 keV range, in agreement with the steep photon index measured in the X-ray band.
This source seems to be peculiar also in the radio waveband; Figure 3 shows the NVSS contours of the source which indicate  that SWIFT J0709.3-1527  is a core dominated radio source
(its 5GHz flux is  558$\pm$31 mJy); however  a structure is evident at either side of the core resembling double  jet emission. 
Of course much more sensitive radio observations are necessary in order to unveil the characteristics of this interesting BL Lac.

\begin{figure}
\includegraphics[angle=-90,width=1.0\linewidth]{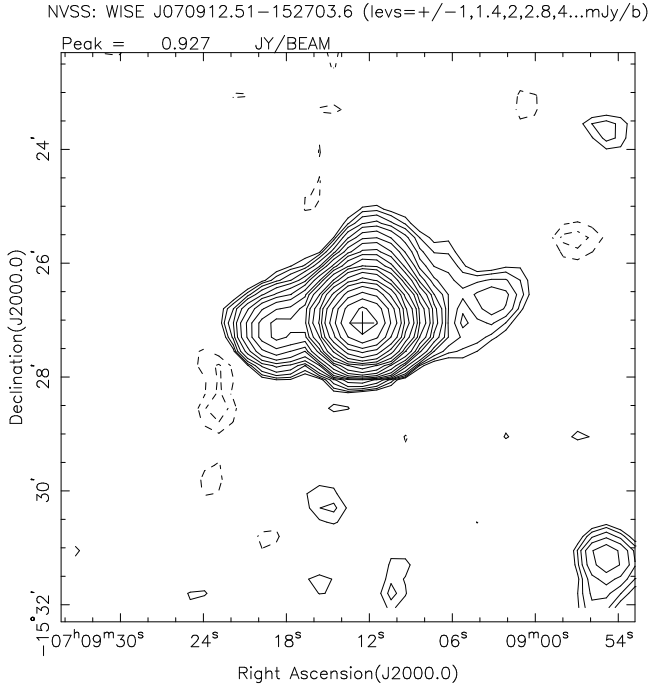}
\caption{NVSS contours of SWIFT J0709.3-1527}
\label{fig3}
\end{figure}

\subsection{NuSTAR data}
\emph{NuSTAR} datasets from both focal plane modules (FPMA and FPMB)  were processed with the NuSTAR--DAS software package (v.1.5.1) 
distributed with HEASOFT version 6.17. 
Event files were calibrated and cleaned with standard filtering criteria using the \texttt{nupipeline} task (version 20150316) of the NuSTAR CALDB.

PKS 0637-752, SWIFT J0845.0-3531, SWIFT J1410.9-4229, MCG-01-40-00, IGR J18308+0928, 4C +21.55 and UGC 12049 are all  well detected in the \emph{NuSTAR} 3-78 keV energy band. 
The FPMA and FPMB spectra of  all targets were extracted from the cleaned event files using a circle of $\sim$ 25-50 arcsec radius, depending on the source brightness, while the background was extracted from nearby circular regions of the same radius. 
The ancillary response files were generated with the  \texttt{numkarf} task, applying corrections for the PSF losses, exposure maps and vignetting. All spectra were binned to ensure a minimum of 30 counts per bin.\\
Also for the \emph{NuSTAR} data a simple power law absorbed by the Galactic column density has been firstly employed including a constant as a free parameter to account for cross-calibration uncertainties between the two telescopes 
and for all  sources this was consistent with unity. Unfortunately \emph{NuSTAR} exposures are quite short and this combined with the relatively weakness of our AGN, does not provide sufficiently high quality spectra to allow
the use of more complex spectral models. 
As mentioned before only the SWIFT J1410.9-4229, MCG-01-40-001, IGR J18308+0928 and UGC 12040, \emph{NuSTAR} data were of  better statistical quality than the \emph{XRT} ones
and therefore the results of this spectral analysis have been considered and reported in Table A1. 
However, for all the 7 sources for which both \emph{NuSTAR} and \emph{XRT} data are available we have been able to compare the two datasets. Furthermore, since \emph{NuSTAR} provides also high energy ($>$20 keV) coverage, 
a comparison between 20-40 keV flux from \emph{NuSTAR} and \emph{INTEGRAL/IBIS} has been considered in order to have also an indication of variability at high energies. Nevertheless it is important to note that \emph{INTEGRAL} fluxes are averaged 
over a long observation period while those of \emph{NuSTAR} are single pointed observations.

\subsubsection{PKS 0637-752} PKS 0637-752, classified in NED as a Blazar (Flat Spectrum Radio Quasar) at z=0.653,  has been detected by \emph{NuSTAR} at 25$\sigma$ signal to noise ratio versus the 65 of the \emph{XRT} observation. 
The two spectra are in good agreement in their shape ($\Gamma_{NuSTAR}$= 1.59$\pm$0.85 and $\Gamma_{XRT}$= 1.75$\pm$0.04) but a significant flux variation is evident when comparing the \emph{NuSTAR} 2-10 keV flux 
of 2.61 $\times$ 10$^{-12}$ erg cm$^{-2}$ s$^{-1}$ with that of \emph{XRT} of 4.72 $\times$ 10$^{-12}$ erg cm$^{-2}$ s$^{-1}$. 
We also found variability in the 20-40 keV band obtaining 3.4 $\times$ 10$^{-11}$ erg cm$^{-2}$ s$^{-1}$ for \emph{NuSTAR} and 5.3 $\times$ 10$^{-12}$ erg cm$^{-2}$ s$^{-1}$
for \emph{IBIS} \citep{bird:2016}. Indeed flux variability is expected for  Blazar objects.

\subsubsection{SWIFT J0845.0-3531}
For SWIFT J0845.0-3531 the two summed \emph{XRT} observations provide higher statistical quality data than the single pointed observation of \emph{NuSTAR}. 
We have analysed both datasets and found no significant flux variation in the soft (2-10) keV  
X-ray band, while changes in the intrinsic absorption have been detected. The best fit of the \emph{XRT} data consists of a power law absorbed by a cold absorption of about 
2 $\times$ 10$^{22}$ cm$^{-2}$ which covers 70$\%$ of the central nucleus. This amount of column density does not vary in the \emph{NuSTAR} observation (N$_{H}$ $\sim$ 2.8 $\times$ 10$^{22}$ cm$^{-2}$) but the data require 
a total (100\% coverage) rather than a partial absorber. 
No excess around 6.4 keV is present in the \emph{NuSTAR} data. Variable and complex absorption is often measured in broad line Seyfert
like  SWIFT J0845.0-3531. In these sources the absorption is likely due to ionized gas located in an accretion disc wind or in the biconical structure associated with
the central nucleus (see \citealt{malizia:2012} and references therein for more details).  Better quality data in the soft X-ray band could provide further insight into this interesting source.

\subsubsection{SWIFT J1410.9-4229} 
SWIFT J1410.9-4229 has recently been  classified as a Seyfert 2 galaxy (Masetti et al. in prep). \emph{NuSTAR} best fit data consists of an absorbed power law with $\Gamma$ $\sim$1.6 and a mild column density of 
N$_{H}$=5.6 $\times$ 10$^{22}$ cm$^{-2}$, while the addition of an iron line does not improve the fit. Comparison in the soft X-rays with \emph{XRT} data and at high energy with \emph{IBIS} ones does not reveal spectral or flux variations.

\subsubsection{MCG-01-40-001}
MCG-01-04-001 is  a Seyfert 2 galaxy whose column density measured by \emph{NuSTAR}  is quite modest (N$_{H}$ $\sim$ 4 $\times$ 10$^{22}$ cm$^{-2}$). 
For this object we have been able to measure an iron line at 6.33 keV with an EW of 247 eV. The comparison of these data with those acquired by \emph{XRT} is important not only because in the 0.3-2 keV range a soft excess has been 
detected, but mostly because a large 2-10 keV flux variation has been found: the \emph{XRT} flux (1.04 $\times$ 10$^{-11}$ erg cm$^{-2}$ s$^{-1}$) is in fact more than twice that of \emph{NuSTAR} (4.3 $\times$ 10$^{-12}$ erg cm$^{-2}$ s$^{-1}$).
The flux variation is also found in the 20-40 keV band since \emph{NuSTAR} measures a hard-X flux of 5.3 $\times$ 10$^{-12}$ erg cm$^{-2}$ s$^{-1}$ while from the last \emph{IBIS} survey we can extract a flux of 2.4 $\times$ 10$^{-11}$ erg cm$^{-2}$ s$^{-1}$.

\subsubsection{IGR J18308+0928} IGR J18308+0928 is a Seyfert 2 galaxy for which the very low statistical quality data of \emph{XRT} ($\sim$ 8$\sigma$) provided us only an estimate of the 2-10 keV flux  ($\sim$ 8  
$\times$ 10$^{-13}$ erg cm$^{-2}$ s$^{-1}$).
The low ratio between 2-10 keV and 20-100 keV fluxes ($\sim$ 0.06) suggests the possibility that this could be a Compton thick AGN \citep{malizia:2011}.
However, the \emph{NuSTAR} observation provides the typical spectral shape of a Compton thin Seyfert  2 object 
absorbed by a column density of 12 $\times$ 10$^{22}$ cm$^{-2}$. The source also shows an iron line at 6.31 keV with an EW of 324 eV. The 2-10 keV flux of 1.3 $\times$ 10$^{-12}$ erg cm$^{-2}$ 
should not be compared with the XRT one due to its low statistics, while
the 20-40 keV fluxes for \emph{NuSTAR} and \emph{IBIS} are in agreement being both around 3 $\times$ 10$^{-12}$ erg cm$^{-2}$ s$^{-1}$.

\subsubsection{4C +21.55} For 4C +21.55 the \emph{XRT} data were of better statistical quality and the results of the  spectral analysis are reported in the table. Furthermore, in the 0.3-2 keV \emph{XRT} range a soft excess 
approximately fitted with a power law  of $\Gamma$=1.97 has been detected. \emph{NuSTAR} data  analysis gives  spectral parameters in agreement with the \emph{Swift/XRT} one; we further checked for the presence of the iron line but it is not required by the data.
However, from  the comparison of  the two datasets  flux variability is found in the 2-10 keV band (F$_{XRT}$ = 1.4 $\times$ 10$^{-11}$ erg cm$^{-2}$ s$^{-1}$ versus F$_{NuSTAR}$ = 7.7 $\times$ 10$^{-12}$ erg cm$^{-2}$ s$^{-1}$)
as well as in the 20-40 keV band (F$_{IBIS}$ = 1.2 $\times$ 10$^{-11}$ erg cm$^{-2}$ s$^{-1}$ versus F$_{NuSTAR}$ = 5.2 $\times$ 10$^{-12}$ erg cm$^{-2}$ s$^{-1}$).\\
4C 21.55 is classified in the \cite{veron:2010} catalogue as a Sey1 and we adopted this classification since not only it is  
a strong radio source (5GHz flux of almost 740 mJy), but also in the NVSS images it appears to be a classical radio galaxy with  extended lobes symmetric with respect to the central source (see Bassani et al. submitted).

\subsubsection{UGC 12040} UGC 12040 has comparable spectra at almost the same S/N ratio (17-18 $\sigma$) in \emph{NuSTAR} and \emph{XRT} but 
the analysis of \emph{NuSTAR} data allowed us to investigate the presence of the iron line. The line is required at $>$90\% significance level, 
is narrow, is located at 6.42$\pm$0.09  keV and has an EW of 646$^{+276}_{-275}$ eV. 
Despite the large errors, the equivalent width is too large for the amount of intrinsic absorption measured (N$_H$ $<$8 $\times$ 10$^{22}$ cm$^{-2}$) although better quality data are needed to assess this observational evidence.
Better data are also necessary  since from the comparison of our spectra
flux variability is found  between \emph{NuSTAR} and \emph{XRT} in the 2-10 keV band (1.5 $\times$ 10$^{-12}$ erg cm$^{-2}$ s$^{-1}$ and 9 $\times$ 10$^{-13}$ erg cm$^{-2}$ s$^{-1}$ respectively) 
and mainly between  \emph{NuSTAR} and \emph{IBIS} in the 20-40 keV range (0.8 $\times$ 10$^{-12}$ erg cm$^{-2}$ s$^{-1}$ and 1.2 $\times$ 10$^{-11}$ erg cm$^{-2}$ s$^{-1}$ respectively).
We note however that the \emph{INTEGRAL} flux has large errors as UGC 12040 was detected at only 4.8$\sigma$ in 20-40 keV band and has only an upper limit in the 40-100 keV band.

\section{Summary and Conclusions}
In this work we report  new AGN detected in the last \emph{INTEGRAL/IBIS} survey \citep{bird:2016} which together with the large AGN sample \citep{malizia:2012} so far detected by \emph{INTEGRAL}, will provide a useful hard
X-ray selected  sample for population  and correlation studies with other wavebans, even though the exposure is not uniform across the sky.
To this aim we have characterized, in terms of optical class and X-ray spectral properties, all the 107 new AGN. The results are presented in this paper for the first time.

This new set of objects consists of 34 broad line or type 1 AGN (Seyfert 1-1.5), 47 narrow line or type 2 AGN (Seyfert 1.8-2) and 18 Blazars. 8 sources are reported as {\it AGN} since their optical class is still undetermined, while
 7  are considered to be good AGN candidates. 
Using  already published X-ray data (55 AGN) or retrieving and analysing   available data from \emph{Swift/XRT}, \emph{Newton-XMM} and \emph{NuSTAR} archives (52 AGN), we have been able 
to study the X-ray spectral characteristics of all 107 new AGN. Among the 7 candidates 
there is only one source (1RXS J112955.1-655542) which has never been observed in the X-ray band. 
Unfortunately the X-ray spectra are not always of  good statistical quality and therefore we cannot employ sophisticated physical models; usually a power law absorbed by the Galactic column density has been 
employed initially  and then the presence of   intrinsic absorption,  a soft excess below 2 keV and  iron line at 6.4 keV, have been investigated. 
Generally we found that the majority of type 2 AGN were intrinsically absorbed while type 1 AGN are generally unobscured as expected in the simplest version of the unified model of AGN.
Nevertheless even with this basic spectral analysis we have been able to identify exceptions and peculiar sources.

Among the 34 type 1 AGN, only 4 were shown to be absorbed by a column density N$_{H}$$>$10$^{22}$ cm$^{-2}$ and these 4 objects (SWIFT J0845.0-3531, IGR J14488-4008, IGR J17488-2338 and IGR J18078+1123)
are all of intermediate class 1.2-1.5 where  X-ray absorption is not so rare \citep{malizia:2012}. 

Almost all of the 47  type 2 galaxies were absorbed by a column density N$_{H}$$>$10$^{22}$ cm$^{-2}$, the only exceptions are: IGR J17379-5957, which is extensively 
discussed in section 3.1.2, and 4 intermediate type 2 AGN.
In fact, among the 11  Seyferts 1.8-1.9 studied in this work, we found 4 galaxies whose X-ray data do not require extra absorption, while  the remaining 
7 are absorbed and one of them (NGC 7479) is even Compton thick. 
As in the case of IGR J17379-5957 (section 3.1.2), the intermediate type 2 objects could be misclassified \citep{trippe:2010}. 
Furthermore, the location of the obscuring material could be another possible explanation for the discrepancy between absorbed and not absorbed intermediate type 2 AGN; 
in fact it is common to found bars or dust lanes in the host galaxies of these AGN.
The study of the correlation between X-ray absorption and optical classification taking into account the environment 
of the AGN is beyond the scope of this work and will be treated  in a forthcoming  paper (Malizia et al in prep.).
It is worth noting that with this update of the \emph{INTEGRAL} AGN catalogue, we find 3 new local Compton thick AGN (NGC 3079, NGC 6926 and NGC 7479), all  already known in the literature but not detected by \emph{INTEGRAL} 
before. We also point out that there is a Seyfert 2, SWIFT J0623.8-3215, which has a large column density  compatible within the errors with a Compton thick AGN (N$_{H}$ =82$^{+34}_{-24}$ $\times$ 10$^{22}$ cm$^{-2}$), 
but the statistical quality of the  \emph{XRT} observation is too poor to draw any conclusion and a better X-ray observation is required to verify this measure.  
Another particular type 2 source that definitely deserves a deeper investigation is the likely Sey 2 IGR J15415-5029. Its \emph{Chandra} observation \citep{tomsick:2012} points to a relatively low column density
(N$_{H}$$<$ 1.1 $\times$ 10$^{22}$ cm$^{-2}$), a negative photon index $\Gamma$=-0.43 and a low 2-10 keV flux ($\sim$ 5 $\times$ 10$^{-13}$ erg cm$^{-2}$ s$^{-1}$). 
Due to the low \emph{Chandra} significance detection (only 46.4 ACIS counts) it could be that the fit gives incorrect values and unfortunately also \emph{XRT} 
has only one short observation detecting the source at a low signal to noise ratio ($<$3$\sigma$).
However, the low flux softness ratio (F$_{(2-10)keV}$/F$_{(20-100)keV}$ = 0.05 see \cite{malizia:2007})  could indicate the presence of a 
much large absorption. Therefore, this source could be considered a Compton thick candidate and its nature  can be clarified only with a good quality X-ray spectrum.

Concerning the 8 objects which are still  optically unclassified, 7 presente mild/strong absorption and so presumably are type 2 AGN, while only one does not have intrinsic absorption and so it is  probably a type 1 AGN.
Two of these objects are strong radio galaxies and indeed in this new set of AGN there are in total 13 radio galaxies which correspond to  12\% of the sample. 
Seven of them are giant radio galaxies (size greater than 0.7 Mpc); in particular B2 1144+35 is
newly detected in the hard X-ray band. The large  percentage of radio galaxies in  hard X-ray selected samples of AGN is extensively discussed in Bassani et al. (submitted)  where it is reported the evidence that the high energy selection seems to favour  
the discovery of giant radio sources. 

In this new set of \emph{INTEGRAL} AGN there are 18 Blazar (FSRQ and BL Lac) of which only 9, half of the sample, have been detected at very high energies by \emph{Fermi}, the remaining are all compact radio sources and definitively deserve 
deeper investigations.

As a final remark we note that in this new set of high energy selected AGN, we found a number of sources which show flux variability.
Seven objects have been  observed both by \emph{NuSTAR} and \emph{XRT} and by comparing the results of the two datasets we found that 5 show 2-10 keV flux variability
and  3  were variable also in 20-40 keV band. 
Finally we draw attention to the intriguing source IGR J17379-5957 for which many \emph{XRT} observations, spanning the period  2008 and 2013, were available and which revealed a quite strong variability on timescales of days/months.

\section*{Acknowledgments}
This research has made use of data obtained from the SIMBAD database operated at CDS, Strasbourg, France; 
the High Energy Astrophysics Science Archive Research Center (HEASARC), provided by NASA's Goddard Space 
Flight Center NASA/IPAC Extragalactic Database (NED). 
The authors acknowledge the ASI financial support via ASI--INAF grant 2013-025-R.0.

\bibliographystyle{mnras}






\appendix
\section{List of the new AGN}
Table A1 lists all the new 107 AGN and, at the end, 7 sources which we consider to be good AGN candidates. 
In column 2 are listed alternative names present in previous catalogues. Coordinates, redshift and class are reported in columns from 3 to 6. In column 7 are listed the satellites used to acquired X-ray data for the 55 sources already known  in the literature or XRT/XMM/NuSTAR net  exposures for data analised in this work. For the sources analised here, XRT detection significance in the 0.3-10 keV band is reported in column 8. 
Intrinsic absorption, photon index and 2-10 keV flux are listed in columns 9, 10, 11 respectively, while the 20-100 keV fluxes extracted from \cite{bird:2016} are reported in column 12.
Finally in column 13,  references for the X-ray spectral parameters are listed.

\clearpage

\begin{landscape}
\begin{table}
\centering
\caption[]{INTEGRAL/IBIS  AGN}
 \begin{tabular}{l l c c  c l l l c l l l  l}
  \hline\hline
 Name	                    & Prev. Cat Name                       &  RA          	 & dec                  &  z        & class & 	exp/instr.	 &  $\sigma$  & 	N$_{H}$\tnote{$^\star$}       &	$\Gamma$	   &   	F$_{S}^{\dagger}$ & F$_{H}^{\ddagger}$    &	REF	\\
 \hline\hline
SWIFT J0034.5-7904	   &  1RXS J003422.2-790525       & 00 34 16.7	 & $-$79 05 20    & 0.07400  & Sy1   & 10500       &    65	  &      -                            & 2.24$^{+0.04}_{-0.04}$ &     6.33               &   1.09           & 1  \\               
Mrk 352	                    &                                                  & 00 59  53.3 	 & $+$31 49 37   & 0.01486 & Sy1   & XMM         &    -           & 	  -  		             & 1.70$^{+0.02}_{-0.01}$  &   11.89	      &   3.21           & 2 	\\
Mrk 1152	                    &                                                  & 01 13  50.1  & $-$14 50 44 	 & 0.05271 & Sy1.5 & 19560\tnote{$^{X}$} &	209 &	-             &  1.61$^{+0.02}_{-0.02}$ &     4.66	      &   2.94           & 1 	\\
Fairall 9	                    &                                                  & 01 23  45.8  & $-$58 48 21 	 & 0.04702	 & Sy1.2 & XMM	&	   -      &            -                       &  1.64$^{+0.03}_{-0.03}$ &	7.40	               &   3.34           & 3	\\
IGR J01528-0845	   & 2MASX J01524845-0843202  & 01 52 48.4   & $-$08 43 21    & 0.03697  & Sy2$^{\S}$	  & 14080        &    6	  & 30.9$^{+73}_{-19}$    & 1.93$^{+1.5}_{-1.5}$       &	0.40	               &   0.85           & 1 \\	
Mrk 584	                    &                                                   & 02 00  26.3 	 & $+$02 40 10   & 0.07877 & Sy1.8 & 9560	&    47	  &      -                             & 2.13$^{+0.07}_{-0.07}$  & 	3.98                 &   1.79           & 1 \\	
IGR J02341+0228	  & QSO B0231+022                      & 02 33  49.0 	 & $+$02 29 25   & 0.32100 & QSO  & XRT	&       -         &       -                            & 2.10$^{+0.30}_{-0.30}$  &	2.90                 &   1.55	    & 4  \\	
SWIFT J0249.1+2627 & 2MASX J02485937+2630391   & 02 48  59.3 	 & $+$26 30 39   & 0.05800 & Sy2    & 8706	& 12   	  & 27.1$^{+10.1}_{- 7.1}$ & 1.7	fixed	                      & 3.73                 &    1.90            & 1	 \\
MCG-02-08-038	  &                                                   & 03 00  04.3    & $-$10 49 29   & 0.03259 & Sy1    & 3961	& 38   	  & 0.36$^{	+0.27}_{-0.27}$ & 1.60$^{+0.20}_{-0.20}$  & 16.7	      &    1.81           & 1	\\
PKS 0312-770	          &                                                    & 03 11  55.2 	 & $-$76 51 51    & 0.22519 & Sy1/QSO & 23562\tnote{$^{X}$}  &	153  & -           & 1.67$^{+0.06}_{-0.06}$  & 2.19	               &    1.21           & 1 \\	
SWIFT J0353.7+3711 & 2MASX J03534246+3714077   & 03 53  42.5 	 & $+$37 14 07   & 0.01865 & Sy2/Liner   & XRT	 &	-          & 3.7$^{+2.0}_{- 1.7}$    & 1.71$^{+0.59}_{-0.55}$  & 3.50                  &    0.89            & 5 \\	
4C +62.08$^{\clubsuit}$ &                                                 & 03 55  40.2 	 & $+$62 40 59   & 1.10900 & Sy1     & 7614	 & 7    	   & -                         	        & 1.47$^{+0.60}_{-0.60}$   & 0.45                &   1.30	     & 1 \\	
SWIFT J0357.6+4153 & 2MASX J03574513+4155049   & 03 57  45.1 	 & $+$41 55 05   &  0.05300 & Sy1.9$^{\S}$   & 10800      & 25 	   & 2$^{+0.7}_{-0.6}$        & 1.61$^{+0.04}_{-0.04}$  & 5.88        &   1.55            & 	1 \\	
SWIFT J0444.1+2813 & 2MASX J04440903+2813003  & 04 44  09.0  & $+$28 13 01   & 0.01127 & Sy2	& XRT        &	   -       & 3.39$^{+0.31}_{-0.25}$ & 1.37$^{+0.11}_{-0.08}$  & 12.34	       &   1.49            & 2 \\	
SWIFT J0450.7-5813  & 2MASX J04514402-5811005    & 04 51  44.0  & $-$58 11 01    & 0.09070 & Sy1.5	& 12000      & 34	   & 0.09$^{+0.04}_{-0.03}$ & 1.45$^{+0.13}_{-0.13}$ & 4.74	       &   1.85            & 1 \\	
MCG -01-13-025	  &                                                    & 04 51  41.5   & $-$03 48 33 	  & 0.01589 & Sy1.2	& XMM	  & -             &      -				 & 1.70$^{+0.03}_{-0.03}$ & 21.3               &    1.47           &	2 \\	
IGR J05081+1722	 & 2MASX J05081967+1721483    & 05 08  19.7 	 & $+$17 21 48   & 0.01750 & Sy2	& XRT        &  -   	   & 2.40$^{+0.30}_{-0.30}$ & 1.80$^{+0.10}_{-0.10}$ & 6.45	       &    1.25           & 6 \\
SWIFT J0515.3+1854 & 2MASX J05151978+1854515    & 05 15  19.8  & $+$18 54 52   & 0.02349 & Sy2	& 7956	  & 8   	   & 11.6$^{+8.6}_{-4.4}$	 & 1.7 fixed                         & 1.91	       &    1.81           & 1 \\	
SWIFT J0516.3+1928 & 2MFGC 04298                            & 05 16  22.7  & $+$19 27 11   & 0.02115 & Sy2	& 7879	  & 10	   &  4.4$^{+2.4}_{-1.3}$      & 1.7 fixed	                 & 1.80                 &    1.53	     & 1 \\	
SWIFT J0544.4+5909 & 2MASX J05442257+5907361    & 05 44  22.6  & $+$59 07 36   & 0.06597 & Sy1.9      & 33800      & 45	   & 1.83$^{+0.3}_{-0.3}$     & 1.62$^{+0.18}_{-0.18}$  & 5.25	        &    2.00          & 1 \\	
IGR J05470+5034	  & 2MASX J05471492+5038251    & 05 47 14.9	 & $+$50 38 25   & 0.03600 & Sy2	& 17500      & 11	   & 15$^{+16}_{-9}$   	 & 1.60$^{+1.60}_{-0.90}$  & 1.40                 &    1.19	      & 1	\\
SWIFT J0623.8-3215	  & ESO 426$-$G 002                      & 06 23  46.4 & $-$32 13 00    & 0.02243 & Sy2	& 31300      & 10	   & 82$^{+34}_{-24}$	          & 1.7 fixed	                 & 1.16	        &    1.94           & 1	 \\                      			
PKS 0637-752	          &                                                     & 06 35  46.5  & $-$75 16 17 	 & 0.65300 & QSO/Blazar & 27100 & 65        &	-                                   & 1.75$^{+0.04}_{-0.04}$  & 4.72	        &    1.66           & 1 \\	
SWIFT J0709.3-1527  & PKS 0706-15                              & 07 09 12.3   & $-$15 27 00   &  0.13900 & BL Lac$^{\S}$  & 606\tnote{$^{X}$}  & 55         & -             & 2.31$^{+0.06}_{-0.06}$  & 13.6              &  $<$1.25     & 1 \\
IGR J07225-3810$^{\clubsuit}$ & PMN J0722-3814	     & 07 22 22.4	 & $-$38 14 55   & 1.023 	  & QSO/Blazar$^{\S}$  &  	XRT	   &   - 	    &    -              			 & 1.52$^{+0.45}_{-0.46}$	 & 0.58	        &    2.08           & 7 \\	
PKS 0723-008	          &                                                      & 07 25  50.6 	 & $-$00 54 57   & 0.12800  & BL Lac  & 29700      & 62	    &		-                   	 & 1.54$^{+0.09}_{-0.11}$	 & 4.97	        &    1.58           & 1 \\	
Mrk 79	                  &                                                       &  07 42 32.8     & $+$49 48 35  & 0.02219	  & Sy1.2   & Suzaku	    &  -	    &       -           			 & 1.51$^{+0.40}_{-0.04}$	 & 15.8              &    4.89	      & 8 \\	
Mrk 1210	                   &                                                      & 08 04 05.8     & $+$05 06 50  & 0.01350	  & Sy2	 & Suzaku	    & 	-           & 33$^{+2.2}_{-2.2}$	 & 1.87$^{+0.23}_{-0.18}$	 & 9.70	        &    5.96          & 9	\\
SWIFT J0845.0-3531  & 1RXS J084521.7-353048            & 08 45 21.4      & $-$35 30 24 	& 0.13700	  & Sy1.2	 & 15400       &	34 	    & 2.4$^{+0.88}_{-0.88}$	 & 1.83$^{+0.31}_{-0.33}$   & 4.64	        &    1.11          & 1 \\	
Mrk 18	                   &                                                      & 09 01 58.4 	& $+$60 09 06 	& 0.01109	  & Sy2	 & XMM        & -             & 18.25$^{3.6}_{-2.7}$	 & 1.62$^{+0.31}_{-0.22}$   & 1.57	        &     2.62         & 2 \\	
1RXS J090431.1-382920 &                                                & 09 04 33.3 & $-$38 29 22  & 0.01603   & Sy1	 & 4657        & 8.4          &   - 				& 1.35$^{+0.38}_{-0.38}$    & 0.92       	&    0.92         & 1 \\	
IGR J09189-4418$^{\clubsuit}$	&  2MASX J09185877-4418302           & 09 18 58.8  	& $-$44 18 30    & 	-          & AGN	 & Chandra   &  -  	   & 4.6$^{+1.1}_{-1.3}$      & 1.4$^{+0.4}_{-0.4}$	          & 2.30	         &   0.53          & 10 \\	
SWIFT J0929.7+6232 & CGCG 312-012                           & 09 29 37.8 	& $+$62 32 39 	 & 0.02561 & Sy2	 & 10310	    & 7	   & 20.5$^{+23.8}_{-12.3}$ & 1.7 fixed	                   & 0.97	         &    2.06         & 1	\\
4C 73.08                      &                                                      & 09 49 45.8 	& $+$73 14 23   & 0.05810 & Sy2	  & XMM	    & - 	   & 92$^{+54}_{-29}$          & 1.7 fixed	                   & 1.81	         &    1.17          & 11	\\
M 81	                           &                                                       & 09 55 33.2 	& $+$69 03 55 	 & 0.00084 & Sy1.8/Liner & 	XMM	 &  -	    &   -  			          & 1.83$^{+0.03}_{-0.02}$    & 11.1	         &    1.11          & 12	 \\
SWIFT J0959.7-3112	 & 2MASX J09594263-3112581       & 09 59 42.6  	& $-$31 12 58 	 & 0.03700  & Sy1	  & 13200       &	69	    &   -      			          & 2.23$^{+0.04}_{-0.04}$    & 7.18	         &     2.21         & 1 \\	
NGC 3079	                  &                                                       & 10 01 57.8 	& $+$55 40 47   & 0.00372  & Sy2	  & BeppoSAX	             & - 	    &1000$^{+1140}_{-530}$ & 1.9 fixed        & 38.0              &     3.72         & 13	 \\
SWIFT J1143.7+7942 & UGC 06728                                 & 11 45 16.0 	& $+$79 40 53 	 & 0.00652	  & Sy1.2	  & XMM	     & -            & 0.01$^{+0.01}_{-0.01}$   & 1.78$^{+0.03}_{- 0.02}$ & 6.51	         &     2.21         & 2 \\	
B2 1144+35B	         &                                                       & 11 47 22.1 	& $+$35 01 08    & 0.06313 & Sy1.9$^{\S}$	  & 4772         &   20       & -                                      & 1.66$^{+0.14}_{-0.14}$  & 3.09              &     1.21         & 1 \\
Mrk 198                      &                                                       & 12 09 14.1  	& $+$47 03 30    & 0.02422 & Sy2	  & XMM         &  -           & 10$^{+0.1}_{-0.1}$          & 1.61$^{+0.04}_{-0.04}$    & 5.12	         &  1.09            & 14 \\	
PKS 1217+02	         &                                                        & 12 20 11.9	& $+$02 03 42	 & 0.24023	  & Sy1.2	  & 14500        &  55	     &	-                                       & 1.99$^{+0.05}_{-0.05}$    & 4.38	         &  1.02            & 1	\\
\hline  
\hline
\end{tabular}
\end{table} 
\end{landscape}

\clearpage

\begin{landscape}
\begin{table}
\contcaption{}
 \begin{tabular}{l l c c  c l l l c l l l  l}
  \hline\hline
Name	                    & Prev. Cat Name                 &  RA 	          &    dec             &  z        & class	& exp/instr	       &  $\sigma$  & 	N$_{H}$$^\star$       &	$\Gamma$	           &   	F$_{S} ^\dagger$ & F$_{H}^\ddagger$    &	REF	\\
 \hline\hline 
PG 1218+305	           &                                                   & 12 21 21.9 	& $+$30 10 37   & 0.18365  & BL Lac   & XMM	      & -            &  -                                     & 2.61$^{+0.02}_{-0.02}$    & 0.26	         &   0.92           & 15 \\	
PG 1222+216	           &                                                  & 12 24 54.4     & $+$21 22 46   & 0.43200	 & QSO/Blazar & XRT    & -            & 	-				   & 2.12$^{+0.14}_{-0.14}$     & 8.30   & 1.64            & 16 \\
IGR J12319-0749$^{\clubsuit}$ & 1RXS J123158.3-074705         & 12 31 57.7     & $-$07 47 18	& 3.66800  & QSO/Blazar & XRT    & -             &  -     			            & 0.96$^{+0.33}_{-0.33}$     & 	5.70	   &  0.98           & 17 \\
Mrk 771	                    &                                                   &  12 32 03.6     & $+$20 09 29 	& 0.06301  & Sy1         & XMM        & -             & 	   -			            & 2.14$^{+0.03}_{-0.02}$     & 	3.01            &  0.83             & 14 \\	
XSS J12303-4232	  & 1RXS J123212.3-421745         & 12 32 11.8    & $-$42 17 52 	& 0.10000    & Sy1.5	   & XRT	     & -  	     &         -    			   & 1.76$^{+0.03}_{-0.04}$      & 5.40             &   1.36            & 18 \\	
SWIFT J1238.6+0928  & 2MASX J12384342+0927362  & 12 38 43.4  & $+$09 27 37 	& 0.08290  & Sy2	 & 16590\tnote{$^{X}$}  & 37 & 32.9$^{+3.7}_{-3.3}$ &	2.3$^{+0.17}_{-0.16}$      & 0.92           &   0.58            & 1 \\		
IGR J13107-5626$^{\clubsuit}$	& 2MASX J13103701-5626551  & 13 10 37.0    & $-$56 26 55     & 	-         & AGN/RG	 & XRT	  &	-             &	39.3$^{+24.4}_{-13.3}$  & 1.8 fixed	                & 1.10	           &  1.02             & 19	\\
IGR J13133-1109	  & PG 1310-108                             & 13 13 05.8    & $-$11 07 42 	& 0.03427	 & Sy1	 & 11400     & 	67	     &   -                                    & 1.87$^{+0.10}_{- 0.11}$     & 8.0	           &  1.34             & 1 \\	
NGC 5100                    &                                                   & 13 20 58.6     & $+$08 58 55 	& 0.03190  & Liner      & XRT	  & - 	              & 14.45$^{+1.41}_{-1.44}$ &	2.02$^{+0.14}_{-0.13}$     & 4.35	   &  1.38             & 14 \\
SWIFT J1344.7+1934  & CGCG 102-048                        & 13 44 15.6  & $+$19 34 00 	& 0.02706  & Sy2/Liner & 13330    & 5	     & 24.6$^{+15}_{-9}$	     & 1.7 fixed	                         & 0.44             &  1.98             & 1  \\	
IGR J13477-4210         & ESO 325-IG 022                      & 13 48 15.2  & $-$42 10 20     & 0.03860  & Sy2       &  5045        & 6              & 6.25$^{+5.3}_{-3.0}$       & 1.7 fixed                             &  1.19           & $<$0.83        & 1 \\
PKS 1355-416	            &                                                  & 13 59 00.2   & $-$41 52 53 	& 0.31300	 & Sy1       & XMM	  &   -		     & 0.27$^{+0.24}_{-0.18}$   &	1.7 fixed	                         & 1.16	   &  0.75             & 20 \\	
SWIFT J1410.9-4229	  & 2MASX J14104482-4228325    & 14 10 44.8   & $-$42 28 33 	& 0.03394  & Sy2$^{\S}$		 & 22230\tnote{$^{N}$}     & 28	     & 5.7$^{+3.4}_{-3.1}$     &	1.57$^{+0.14}_{-0.13}$ 	          & 2.07	   &  1.02              & 1 \\	
SWIFT J1417.7+2539  & 1RXS J141756.8+254329        & 14 17 56.7 & $+$25 43 26 	& 0.23700  & BL Lac	 & 14600     & 84	     &   -                           	     & 2.07$^{+0.03}_{-0.03}$      & 12.0	            &  2.17             & 1 \\	
NGC 5674	                    &                                                   & 14 33 52.2 & $+$05 27 30 	& 0.02493  & Sy1.9	 & XRT	  & -               & 11.2$^{+6.1}_{-2.7}$	     & 2.14$^{+0.06}_{-0.25}$      & 4.85	   &  1.81             & 14 \\ 	
SWIFT J1436.8-1615	  & 2MASX J14364961-1613410    & 14 36 49.6 &	$-$16 13 41 	& 0.14454  & Sey1/QSO	 & 9400	   & 45	     &   -                                     &	1.72$^{+0.06}_{-0.06}$      & 7.19	   &  2.00              & 1	\\
IGR J14488-4008$^{\clubsuit}$	& WISE J144850.99-400845.6    & 14 48 51.0 &	$-$40 08 46	& 0.12300  & Sy1.2	 & XMM	   & - 		     & 7.85$^{+3.04}_{-2.37}$   & 1.71$^{+0.16}_{-0.17}$      & 5.27	    & 0.75              & 21	\\
IGR J14492-5535$^{\clubsuit}$	&  CXO J144917.3-553544         & 14 49 17.3 & $-$55 35 45	&	-	 & AGN	 & Chandra  & -               &	12		                       & 1.8 fixed	                          & 3.10           & 1.81	           & 22 \\	
PKS 1451-375            &                                                      & 14 54 27.4 & $-$37 47 33 	& 0.31405  & Sy1.2	 & 1643         &   12.5      &  - 			              & 1.74$^{+0.27}_{-0.27}$       & 3.18	    & 0.98              & 1 \\	
1RXS J150101.7+223812 &                                               & 15 01 01.8 & $+$22 38 06 & 0.23500 & BL Lac & BeppoSAX & 	-             & -             			     &	2.62$^{+0.23}_{-0.12}$	&0.70            &  3.69             & 23  \\	
SWIFT J1508.6-4953 & 1RXS J150839.0-495304            & 15 08  39.0 &	 $-$49 53 02      &        -         & Blazar  & XRT           &	-            &  -				     & 1.41$^{+0.10}_{-0.10}$       & 2.80           &  0.96             & 24	\\
PKS 1510-089          &                                                        & 15 12 50.5  &	 $-$09 06 00 	& 0.36000	  & QSO/Blazar & Suzaku & -         &  -				     & 1.37$^{+0.01}_{-0.01}$       & 6.31	    &  3.85             & 25	\\
ESO 328-36	          &                                                      & 15 14 47.2  & $-$40 21 35      & 0.02370   & Sy1.8	 & 10500        &	15	     & -                        		     & 1.77$^{+0.21}_{-0.21}$       & 7.13	    &  0.85             &  1 	\\
IGR J15301-3840	 & ESO 329- G 007                          & 15 30 08.0  & $-$38 39 07      & 0.01553   & Sy2$^{\S}$	       &  3099          & 9.8        & 2.27$^{+1.96}_{-1.36}$   & 1.74$^{+1.30}_{-1.07}$       & 3.34           & 	0.77             &  1 \\		  	
MCG-01-40-001          &                                                     & 15 33 20.7  & $-$08 42 02      & 0.02271  & Sy2	 & 21640\tnote{$^{N}$}      & 45          & 3.9$^{+1.95}_{-1.83}$    & 	1.54$^{+0.09}_{-0.08}$	& 4.27	    &  1.77             & 1 \\	
IGR J15359-5750$^{\clubsuit}$ & CXO J153602.7-574853         & 15 36 02.8   & $-$57 48 53      &  	-	  & AGN	 & XMM	      &  -	      & 20.1$^{+2.5}_{-2.9}$      &	1.85$^{+0.27}_{-0.25}$	& 4.97	    &  2.06              & 26 \\
IGR J15415-5029$^{\clubsuit}$ & WKK 5204 	            & 15 41 26.4   & $-$50 28 23	 & 0.03200	  & Sy2?	 & Chandra     & -             & $<$1.1		              & -0.43$^{+0.68}_{-0.44}$      & 0.54	    &  1.04             & 10 \\	
IGR J16058-7253 (1)	 & LEDA 259580                            & 16 06 06.9   & $-$72 52 42	 & 0.09000  & Sy2  	 &  25200        & 8 	      &	 37.9$^{+10.9}_{-15.8}$  &	1.75$^{+1.55}_{-1.64}$	 & 3.18	    &   0.90            & 1	\\
IGR J16058-7253 (2)  & LEDA 259433                            & 16 05 23.2   & $-$72 53 56       & 0.06900  & Sy2?	 &  25200        & 15	      & 17.4$^{+6}_{-5}$             & 1.90$^{+0.87}_{-0.49}$       & 3.50          &   0.60            & 1 \\		
Mrk 1498                  &                                                       & 16 28 04.0   & $+$51 46 31 	 & 0.05470  & Sy1.9	 & Suzaku	      & - 	      &	 58$^{+11}_{-11}$            & 1.80$^{+0.03}_{-0.03}$       & 9.00          &   5.09            & 27  \\	
SWIFT J1630.5+3925 & 2MASX J16303265+3923031   & 16 30 32.6  & $+$39 23 03 	 & 0.03056  & Sy2	 & XMM	      & - 	      & 40$^{+5}_{-7}$                & 1.20$^{+0.3}_{-0.4}$	          & 1.25	     &  1.59             & 29  \\	
IGR J17111+3910	 & 2MASX J17110531+3908488    & 17 11 05.3  & $+$39 08 49      & 	-	  & AGN	 & 4922           & 16         & -                                        & 1.96$^{+0.17}_{-0.17}$	& 1.47   & 	 0.96             & 1	\\
IGR J17204-3554$^{\clubsuit}$	 &                                    & 17 20 21.6  & $-$35 54 32      &     -           & AGN	 & XRT	      & - 	      &	 16$^{+3}_{-3}$                & 1.6$^{+0.4}_{-0.4}$             & 2.50              &  1.15            & 28  \\	
Mrk 506	                   &                                                     & 17 22 39.9  & $+$30 52 53 	 & 0.04303  & Sy1.5	 & ASCA	      & -             & $<$0.05                           & 1.93$^{+0.07}_{-0.05}$     	 & 6.30	      &   1.81           &  30	\\
SWIFT J1723.5+3630 & 2MASX J17232321+3630097    & 17 23 23.2  & $+$36 30 10 	 & 0.04000	  & Sy1.5    & 8745	      & 40          & 0.19$^{+0.06}_{-0.05}$    & 1.56$^{+0.11}_{-0.11}$       & 8.39	      &   1.79           & 1 \\	
PKS 1730-13	           &                                                    & 17 33 02.7 & $-$13 04 50 	 & 0.90200	  & QSO/Blazar & 40800 &	34           &	 0.87$^{+0.09}_{-0.08}$   & 1.61$^{+0.15}_{-0.15}$       & 1.46	      &    1.30          & 1 	\\
\hline
 \hline
\end{tabular}
\end{table} 
\end{landscape}

\clearpage

\begin{landscape}
\begin{table}
\contcaption{}
 \begin{tabular}{l l c c  c l l l c l l l  l}
  \hline\hline
Name	                          & Prev. Cat Name                        &  RA  	          &    dec        &  z        & class	& 	exp/instr	       &  $\sigma$  & 	N$_{H}$$^\star$                   &	$\Gamma$	           &   	F$_{S} ^{(\dagger)}$ & F$_{H}^{(\ddagger)}$    &	REF	\\
 \hline\hline
IGR J17348-2045$^{\clubsuit}$  & NVSS J173459-204533   & 17 34 59.1 & $-$20 45 34 &   0.044             &  Sy2$^{\S}$      & XMM              & -              & 17$^{+7}_{-5}$              & 1.5$^{+0.8}_{-0.7}$        & 2.2                                & 1.7               & 31\\
IGR J17379-5957	  & ESO 139- G 012                              & 17 37 39.1  & $-$59 56 27      & 0.01702  &  Sy2	  & 17400        & 61	       &  -                                      & 1.80$^{+0.07}_{-0.07}$	  & 6.02	      &  1.17             & 1 \\	
PKS 1741-03	           &                                                          &  17 43 58.8  & $-$03 50 05  	 & 1.05400  & QSO/Blazar	 & 19800 & 25          & -                                       & 1.35$^{+0.19}_{-0.18}$       & 2.09	      &   0.70           & 1  \\	
IGR J17448-3232$^{\clubsuit}$ & CXOU J174453.4-323254 & 17 44 55.0  & $-$32 32 00      & 0.05500  &  Cluster/Blazar & XMM & -           & 2.51$^{+0.25}_{-0.23}$   & 1.31$^{+0.09}_{-0.09}$       & 1.65            & 1.11             &  32\\
IGR J17488-2338$^{\clubsuit}$	&                                            & 17 48 39.0  & $-$23 35 21	 & 0.24000  & Sy1.5	  & XMM	       &  -            & 1.14$^{+0.26}_{-0.23}$   & 1.37$^{+0.11}_{-0.11}$       & 2.0             &   1.58            & 33 \\	
IGR J17520-6018$^{\clubsuit}$	 & 2MASX J17515581-6019430 & 17 51 55.8  & $-$60 19 43	 & 0.11200  & Sy2	  & XRT	       & -	       &	13		                & 1.8	fixed	                             & 2.55	      &   2.19             & 34	\\
NGC 6552	                   &                                                            & 18 00  07.2 & $+$66 36 54 	 & 0.02649  & Sy2	  & XMM	       & -             & 71$^{+40}_{-10}$            & 2.85$^{+0.37}_{-0.13}$	   & 0.43	      &    0.85                 &  35	\\
IGR J18078+1123	  & 2MASX J18074992+1120494           & 18 07 49.9 &	 $+$11 20 49	 & 0.07800  & Sy1.2 & 17200     & 52           & 1.79$^{+0.47}_{-0.42}$    & 2.00$^{+0.19}_{-0.20}$        & 6.32	      &    1.26                 & 1	\\
IGR J18129-0649$^{\clubsuit}$ & PMN J1812-0648	            & 18 12 50.9 & $-$06 48 24       & 0.77500	  & Sy1/QSO$^{\S}$ 	   & 	XRT	       &   -           &    -                                    & 1.44$^{+0.41}_{-0.43}$        & 2.74	      &    0.66                & 7	 \\
SWIFT J1821.6+5953 & 2MASX J18212680+5955209           & 18 21 26.8  & $+$59 55 21	 & 0.0990  & Sy2	   & 10500        & 11          & 7.65$^{+3.7}_{-2.3}$        & 1.7	fixed                              & 1.49	      &    1.41                 &  1	\\
IGR J18308+0928$^{\clubsuit}$	  & 2MASX J18305065+0928414 & 18 30 50.6  & $+$09 28 42     & 0.01900   & Sy2	   & 22720\tnote{$^{N}$}        & 24	       & 11.9$^{+7.8}_{-6.5}$        & 1.54$^{+0.27}_{-0.24}$       & 1.30	     &      1.32                & 1	\\
Fairall 49	                   &                                                            & 18 36  58.3 & $-$59 24 09 	 & 0.02002   & Sy2	   & XMM	       &  -            & 1.24$^{+0.08}_{-0.11}$	& 2.06$^{+0.03}_{-0.03}$      & 12.0	     &       1.51                & 36	\\
SWIFT J1845.4+7211 & CGCG 341-006                                 & 18 45  26.2 & $+$72 11 02 	 & 0.04630   & Sy2       & 11290         & 18.5       & 3.24$^{+0.57}_{-0.46}$     & 1.7 fixed                            & 4.22	      &    1.00                  & 1	\\
PBC J1850.7-1658      & PMN J1850-1655                              & 18 50 51.6  & $-$16 55 58      & -               & AGN/RG        & 4295           & 10.5       & 1.49$^{+0.62}_{-0.44}$     & 1.7 fixed                             & 2.19         &    1.28                   & 1\\
IGR J18538-0102$^{\clubsuit}$	& 2XMM J185348.4-010229   & 18 53 48.4  & $-$01 02 30	 & 0.14500	   & Sy1	   & XMM	       &  -            & 0.41$^{+0.1}_{-0.1}$	& 1.56$^{+0.08}_{-0.08}$      & 4.0	               &    1.02                 & 26 \\	
SWIFT J1933.9+3258 & 2MASS J19334715+3254259             & 19 33  47.1 & $+$32 54 26 	 & 0.0565    & Sy1.2	   & XRT	       & - 	       & $<$0.04                             & 2.08$^{+0.05}_{-0.04}$      & 13.8	      &     1.30                 & 18	\\
IGR J19443+2117$^{\clubsuit}$ & 2MASS J19435624+2118233	            & 19 43 56.2  & $+$21 18 23	 &	-	   & BL Lac? & XRT	       & -             & 0.54$^{+0.12}_{-0.13}$	& 2.04$^{+0.12}_{-0.12}$	   & 18.3	      &     1.00                 & 37	\\
1ES 1959+650	           &                                                              & 19 59  59.8 & $+$65 08 55 	 & 0.04700   & BL Lac   & XRT	       &   -           & 0.07$^{+0.04}_{-0.04}$	& 1.93$^{+0.11}_{- 0.10}$     & 91.7 	      &     1.57                 & 2	\\
SWIFT J2018.4-5539	  & PKS 2014-55                                       &  20 18  01.3 & $-$55 39 31      & 0.06063   & Sy2	    & 10800       & 13           & 31.4$^{+11.6}_{-9}$ 	         & 1.7 fixed	            & 4.01	      &     3.64                 & 1	\\
IGR J20159+3713	  & LEDA 101342                                      & 20 15 28.8  & $+$37 11 00     & 0.857       & QSO/Blazar$^{\S,a}$ &  XRT &  -             & 0.78$^{+0.33}_{-0.29}$     & 1.66$^{+0.22}_{-0.20}$      & 2-5 (var)   &     1.11                 & 38 \\	
NGC 6926	                   &                                                               & 20 33  06.1  & $-$02 01 39 	 & 0.01961  & Sy2	     & XMM       & -              & 100                                     & 1.7 fixed                            & 0.07	       &     1.36                &  39 \\
4C +21.55	                   &                                                               & 20 33 32.0   & $+$21 46 22 	 & 0.17350	  & Sey1        & 19800 &	75            & 0.60$^{+0.42}_{-0.20}$      & 1.68$^{+0.50}_{-0.11}$      & 14.0	       &    2.91                 & 1	\\
IGR J21178+5139$^{\clubsuit}$  & 2MASX J21174741+5138523     	  & 21 17 47.2   & $+$51 38 54 	 &     -           & AGN		   & XRT	     & -      & 2.11$^{+1.52}_{-1.03}$	& 1.8	 fixed                             & 2.1	       &    1.81                 & 40 \\
CTS 109	                   &                                                               & 21 32  02.1 & $-$33 42 54       & 0.02997  & Sy1.2	     & XMM	   &     -            & -        				& 1.55$^{+0.02}_{-0.02}$      & 7.6	       &    1.98                  & 41	\\
SWIFT J2156.2+1724 & 2MASX J21561518+1722525             & 21 56 15.2  & $+$17 22 53      &  0.03400   & Sy1.8$^{\S}$	        & 8456	   &     9	       & 21.5$^{+37.5}_{-9.83}$	& 1.7 fixed                             & 1.77	       &    2.00                  & 1	\\
UGC 12040	          &                                                               & 22 27  05.8 & $+$36 21 42 	 & 0.02133  & Sy1.9	     & 21340\tnote{$^{N}$} & 17	  & $<$ 8.3                  & 1.63$^{+0.29}_{-0.19}$	   & 0.86	       &    2.62                  & 1	\\
KAZ 320	                   &                                                              &  22 59  32.9 & $+$24 55 06 	 & 0.03450	  & NLSy1	     & 7305         & 52 	      & - 			                  & 1.91$^{+0.05}_{-0.05}$      & 10.1           &    1.49                 & 1	\\
NGC 7479	                   &                                                              & 23 04  56.6 & $+$12 19 22 	 & 0.00794  & Sy1.9	     & XMM		& -          & 201$^{+493}_{-122}$	         & 1.9 fixed	                     & 0.33	       &    1.40                 & 13	\\
RHS 61	                   &                                                             &  23 25  54.2 & $+$21 53 14 	 & 0.12000	  & Sy1	     & 3414         & 24      &  - 			                  & 1.86$^{+0.11}_{-0.11}$      & 4.32	       &    1.43                 & 1	\\
PKS 2325+093	          &                                                              & 23 27  33.6 &	 $+$09 40 09 	 & 1.84300  & QSO/Blazar &  XRT    & -           & - 					 & 1.11$^{+0.17}_{-0.17}$      & 2.9	       &     2.45                & 42	\\
IGR J23558-1047$^{\clubsuit}$& 1WGA J2355.9-1045	 & 23 55 59.3   & $-$10 46 45	 & 1.10800  & Sy1/QSO	 & XRT    & -           & - 					 & 1.97$^{+0.80}_{-0.69}$      & 0.08	       &     5.26                & 43	\\	
\hline
\hline
\end{tabular}
\end{table} 
\end{landscape}
\clearpage

\begin{landscape}
\begin{table}
\contcaption{}
 \begin{tabular}{l l c c  c l l l c l l l  l}
 \multicolumn{13}{c}{{\bf AGN CANDIDATES}} \\
  \hline\hline
Name	                       & Prev. Cat Name   &  RA  	          &    dec        &  z        & class	& 	exp/instr	       &  $\sigma$  & 	N$_{H}$$^\star$                   &	$\Gamma$	           &   	F$_{S} ^{(\dagger)}$ & F$_{H}^{(\ddagger)}$    &	REF	\\
 
\hline
\hline
IGR J02447+7046$^{\clubsuit}$\tnote{b}  & NVSS J024443+704946  & 02 44 23.7 & $+$70 29 05  &   -            & cand       & 4488               & 4.4            &       -                             & 1.7 fixed                         & 0.18                              &  0.96            & 1 \\	
IGR J02447+7046$^{\clubsuit}$\tnote{b}  &                                          & 02 43 43.0 & $+$70 50 38  & 0.30600  & Sy 1.2     & 4488               & 4	       &        -       		            & 1.7 fixed                         & 0.13                               &  0.96            & 1 \\	
RXS J112955.1-655542                            &                                           & 11 29 55.1 & $-$65 55 42 &    -           & cand        & -                     &  -              & -                                    & -                                      &         -                             &   0.94          & 44\\
IGR J14466-3352$^{\clubsuit}$                & NVSS J144637-335234    & 14 46 37.4 & $-$33 52 34 &   -            & cand        &  XRT              &                 &  -                                   & 1.8  fixed                         & 0.07                              & 0.79             & 45 \\
IGR J16413-4046$^{\clubsuit}$                & CXOU J164119.4-404737 & 16 41 19.5 & $-$40 47 38 &   -            & cand        &  XRT              & -               & 10                                 & 1.8 fixed                          & 1.40                              & 1.45             & 46 \\
IGR J16560-4958$^{\clubsuit}$              & CXOU J165551.9-495732  & 16 55 51.9 & $-$49 57 32 &  -             & cand        & Chandra        & -               & 2.3$^{+0.7}_{-0.4}$       & 2.2$^{+0.5}_{-0.4}$        & 17.0                                 & 1.10             & 47 \\
AX J183039-1002$^{\clubsuit}$              &                                             & 18 30 38.3 & $-$10 00 25 &   -            &  cand       & Chandra         & -              &11.4$^{+3.8}_{-2.9}$     & 1.01$^{+0.57}_{-0.38}$   & 3.3                               & 0.73             & 48\\
IGR J20569+4940$^{\clubsuit}$             & MG4 J205647+4938            & 20 56 42.7 & $+$49 40 07 &  -             & cand       & XRT               &  -             & 0.53$^{+0.18}_{-0.16}$ & 2.32$^{+0.18}_{-0.17}$  & 1.19                            & 0.83             & 49 \\
\hline
\hline
\end{tabular}
\end{table} 

\begin{tablenotes}
      \item  $^{(\star)}$ : intrinsic column density in unit of 10$^{22}$ cm$^{-2}$; $^{(\dagger)}$: 2-10 keV flux in units of 10$^{-12}$ erg cm$^{-2}$ s$^{-1}$; $^{(\ddagger)}$: 20-100 keV flux in units of 10$^{-11}$ erg cm$^{-2}$ s$^{-1}$.
       $^{(X)}$: XMM data; $^{(N)}$: NuSTAR data,  $^{(\S)}$: new classification in Masetti et al. in prep.;   [a]: the blazar object is blended with a CV but \cite{bassani:2014} have estimated that the 90\% hard X-ray fluxed comes from the AGN;   
                                              [b]: two X-ray sources, 1 identified as a Sy 1.2 and 1 AGN candidate, are present in the IBIS error box;    

                                    References: (1): {\bf spectral analysis performed in this work}; (2): \citealt{winter:2009}; (3) \citealt{tombesi:2010}; (4): \citealt{ricci:2012}; (5) \citealt{parisi:2012}; (6)\citealt{ballo:2015};
                                    (7) \citealt{molina:2012}; (8) \citealt{winter:2012}; 
                                    (9): \citealt{matt:2009};  (10): \citealt{tomsick:2012}; (11): \citealt{evans:2008}; (12): \citealt{brightman:2011}; (13): \citealt{Iyomoto:2001};
                                    (14): \citealt{vasudevan:2013}; (15): \citealt{blustin:2004}; (16): \citealt{tavecchio:2011};  (17): \citealt{bassani:2012};
                                    (18): \citealt{landi:2007}; (19) \citealt{landi:2010}; (20) \citealt{mingo:2014}; (21): \citealt{molina:2015}; (22): \citealt{sazonov:2008}; (23): \citealt{donato:2005}; (24): \citealt{landi:2013}; 
                                    (25): \citealt{abdo:2010}; (26): \citealt{malizia:2010}; (27): \citealt{eguchi:2009}; (28): \citealt{castangia:2013}; (29): \citealt{bassani:2005}; 
                                    (30): \citealt{shinozaki:2006}; (31): \citealt{bozzo:2012}; 
                                    (32): \citealt{barriere:2015};
                                    (33) \citealt{molina:2014}; (34): \citealt{maiorano:2011}; (35): \citealt{shu:2007}; (36): \citealt{iwasawa:2004}; (37): \citealt{landi:2009};
                                    (38): \citealt{bassani:2014};  (39) \citealt{greenhill:2008};
                                    (40): \citealt{malizia:2007}; (41): \citealt{cardaci:2011}; (42): \citealt{ghisellini:2009}; (43): \citealt{malizia:2011}; (44): \citealt{edelson:2012}; (45): \citealt{Landi:2010b}; 
                                    (46):\citealt{landi:2011b}; (47): \citealt{tomsick:2012}; (48): \citealt{bassani:2009}; 
                                    (49): \citealt{landi:2010}.

\end{tablenotes}
\end{landscape}


\bsp	
\label{lastpage}

\end{document}